\documentclass[twocolumn,showpacs,preprintnumbers]{revtex4}
\usepackage{amsmath}
\usepackage{dcolumn}
\usepackage{bm}
\usepackage{graphicx}
\usepackage{amsfonts}
\usepackage{amssymb}
\usepackage{mathrsfs}
\usepackage{color}
\usepackage{braket}

\begin{document}

\title{No spin-localization phase transition in the spin-boson model without local field}
\author{Tao Liu$^{1,2}$}
\email{liutao849@163.com}
\author{Mang Feng$^{2}$}
\email{mangfeng@wipm.ac.cn}
\author{Lei Li$^{1}$}
\author{Wanli Yang$^{2}$}
\author{Kelin Wang$^{3}$}
\affiliation{$^{1}$ The School of Science, Southwest University of Science and
Technology, Mianyang 621010, China \\
$^{2}$ State Key Laboratory of Magnetic Resonance and Atomic and
Molecular Physics, Wuhan Institute of Physics and Mathematics,
Chinese Academy of Sciences, and Wuhan National
Laboratory for Optoelectronics, Wuhan, 430071, China \\
$^{3}$ The Department of Modern Physics, University of Science and
Technology of China, Hefei 230026, China}

\begin{abstract}
We explore the spin-boson model in a special case, i.e., with zero
local field. In contrast to previous studies, we find no possibility
for quantum phase transition (QPT) happening between the localized
and delocalized phases, and the behavior of the model can be fully
characterized by the even or odd parity as well as the parity
breaking, instead of the QPT, owned by the ground state of the
system. Our analytical treatment about the eigensolution of the
ground state of the model presents for the first time a rigorous
proof of no-degeneracy for the ground state of the model, which is
independent of the bath type, the degrees of freedom of the bath and
the calculation precision. We argue that the QPT mentioned
previously appears due to incorrect employment of the ground state of
the model and/or unreasonable treatment of the infrared divergence existing in the spectral
functions for Ohmic and sub-Ohmic dissipations.
\end{abstract}

\maketitle

A two-level system coupled to an environment provides a unique
test-bed for fundamental interests of quantum physics. Denoting the
environment by a multi-mode harmonic oscillator, the spin-boson
model (SBM) \cite{weiss, leggett} presents a phenomenological
description of the open quantum system, which plays an important
role in quantum information science and condensed matter physics.
Particularly, for the SBM at zero temperature, it has attracted
intensive interests for the quantum phase transition (QPT) happening
between localized and delocalized phases regarding the spin.

The standard SBM Hamiltonian in units of $\hbar=1$ is given by
\cite{leggett},
\begin{equation}
H=\frac {\epsilon}{2}\sigma_{z}-\frac {\Delta}{2}\sigma_{x}+
\sum_{k}\omega_{k}a^{\dagger}_{k}a_{k} +
\sum_{k}\lambda_{k}(a^{\dagger}_{k} + a_{k})\sigma_{z},
\end{equation}
where $\sigma_{z}$ and $\sigma_{x}$ are usual Pauli operators,
$\epsilon$ and $\Delta$ are, respectively, the local field (also
called c-number bias \cite {leggett}) and tunneling regarding the
two levels of the spin. $a^{\dagger}_{k}$ and $a_{k}$ are creation
and annihilation operators of the bath modes with frequencies
$\omega_{k}$, and $\lambda_{k}$ is the coupling between the spin and
the bath modes. The effect of the harmonic oscillator environment is
reflected by the spectral function
$J(\omega)=\pi\sum_{k}\lambda_{k}^{2}\delta (\omega-\omega_{k})$ for
$0<\omega <\omega_{c}$ with the cutoff energy $\omega_{c}$. In the
infrared limit, i.e., $\omega\rightarrow$0, the power laws regarding
$J(\omega)$ are of particular importance. Considering the low-energy
details of the spectrum, we have
$J(\omega)=2\pi\alpha\omega_{c}^{1-s}\omega^{s}$ with $0<\omega
<\omega_{c}$ and the dissipation strength $\alpha$. The exponent $s$
is responsible for different bath with super-Ohmic bath $s>$1, Ohmic
bath $s =$1 and sub-Ohmic bath $s <$1.

The local field $\epsilon$ introduces asymmetry in the model, which
was considered to be less important than the tunneling $\Delta$ and
thereby sometimes neglected for convenience of treatments. For the
Ohmic dissipation, it was mentioned \cite {hur1,Bu1} that the SBM
has a delocalized and a localized zero temperature phase, separated
by a Kosterlitz-Thouless (KT) transition in the case of
$\epsilon=0$. In the delocalized phase, realized at small
dissipation strength $\alpha$, the non-degenerate ground state
behaves like a damped tunneling particle. In contrast, for large
$\alpha$, the dissipation leads to a localization of the particle in
one of the two $\sigma_{z}$ eigenstates, implying a doubly
degenerate ground state. On the other hand, intensive studies have
recently been paid on the sub-Ohmic dissipation, which also
demonstrated the QPT between localized and delocalized phases \cite
{Bu1,hur1,kehrein,vojta2,vojta1, hur,
winter,Bu2,Bu3,and,rmp1,cheng,AA,
chin,vojta3,wong,florens,zheng1,vojta4,zheng2}. In particular, the
QPT could also happen in the absence of the local field
($\epsilon=0$) \cite {vojta1,hur,winter} from a non-degenerate
ground state with zero magnetization below a critical coupling to a
twofold-degenerate ground state with finite magnetization for a
coupling larger than the critical value. The QPT has been found so
far by different numerical studies, such as numerical
renormalization group (NRG) technique \cite
{vojta2,vojta1,hur,Bu1,and,rmp1,Bu2,Bu3,cheng}, quantum Monte Carlo
\cite{winter}, the exact diagonalization \cite{AA}, the density matrix
renormalization group approach \cite{wong}, and variational matrix
product state approach \cite{vojta4}. Analytical studies, such as
unitary transformation method \cite{zheng1,zheng2}, were also
employed. The latest work by variational method also showed the
appearance of the QPT between delocalization and localization
\cite{chin} in the case of $\epsilon=0$.

Concentrating on the SBM without the local field, we argue in the
present work no possibility with the QPT happening between the
localized and delocalized phases in above dissipative cases,
opposite to previous results. The main behavior of the model could
be characterized by the parity keeping or breaking, rather than the
QPT, owned by the ground state in the variation of some
characteristic parameters. The remarkable feature of our treatment
is the analytical investigation, which is independent of the bath
type, the degrees of freedom of the bath and the calculation
precision. The key step of our treatment is the decomposition of the
special SBM Hamiltonians in such case into two decoupled
sub-Hamiltonians, and the key point is the focus on the parity of
the ground state of the model. We show that the significant changes
regarding the ground state of the model, if can be called 'QPT',
happen only when the parity breaks down, corresponding to appearance
of the local field. Our investigation of the magnetization
analytically indicates that the QPT discussed previously by
different numerical studies is due to unreasonable treatment of the
ground state and/or of the infrared divergence inherently existing in
the spectral functions for Ohmic and sub-Ohmic dissipations.

We get started from the special SBM with zero local field. In such a
case, we denote the Hamiltonian $H$ by $H_{SB}$ with
\begin{equation}
H_{SB}=-\frac {\Delta}{2}\sigma_{x}+
\sum_{k}\omega_{k}a^{\dagger}_{k}a_{k} + \sum_{k}\lambda_{k}(a^{\dagger}_{k}
+ a_{k})\sigma_{z},
\end{equation}
and introduce a parity operator
\begin{equation}
\Pi=\sigma_{x}e^{i\pi\sum_{k}a^{\dagger}_{k}a_{k}},
\end{equation}
to commute with $H_{SB}$, i.e., $[H_{SB}, \Pi]=0$. Moreover, by
introducing a unitary transformation
\begin{equation}
U=%
\begin{pmatrix}
1 & e^{-i\pi\sum_{k}a^{\dagger}_{k}a_{k}} \\
-e^{i\pi\sum_{k}a^{\dagger}_{k}a_{k}} & 1%
\end{pmatrix}
/\sqrt{2},   \notag
\end{equation}
we may diagonalize $H_{SB}$ to be two decoupled sub-Hamiltonians,
\begin{equation}
H_{d}= U H_{SB} U^{+} =
\begin{pmatrix}
H^{+} & 0 \\
0 & H^{-}%
\end{pmatrix}
,
\end{equation}
where
\begin{equation}
\begin{split}
\label{eq:H+-}
H^{+}\! &= -\frac {\Delta}{2} e^{i\pi\sum_{k}a^{\dagger}_{k}a_{k}} +
\sum_{k}[\omega_{k}a^{\dagger}_{k}a_{k} + \lambda_{k}(a^{\dagger}_{k} +
a_{k})], \\
H^{-}\! &=  \frac {\Delta}{2}e^{i\pi\sum_{k}a^{\dagger}_{k}a_{k}} +
\sum_{k}[\omega_{k}a^{\dagger}_{k}a_{k} - \lambda_{k}(a^{\dagger}_{k} +
a_{k})].
\end{split}
\end{equation}
It implies that the eigensolution of $H_{d}$ consists of two
independent eigensolutions of $H^{+}$ and $H^{-}$. Since no local
unitary transformation can change the eigenvalues, the ground state
of $H_{SB}$ should be the lower one of the ground states of $H^{+}$
and $H^{-}$.

For the eigensolutions $H_{d}\begin{pmatrix}
|\varphi^{+}\rangle \\
0%
\end{pmatrix}=E^{+}\begin{pmatrix}
|\varphi^{+}\rangle \\
0%
\end{pmatrix}$ and
$H_{d}\begin{pmatrix}
0 \\
|\varphi^{-}\rangle
\end{pmatrix}=E^{-}\begin{pmatrix}
0 \\
|\varphi^{-}\rangle
\end{pmatrix}$,
it is easy to prove that the eigenfunctions are, respectively, of
even-parity and odd-parity with respect to the parity operator
$\sigma_{z}$ since $\sigma_{z}
\begin{pmatrix}
|\varphi^{+}\rangle \\
0%
\end{pmatrix}
=
\begin{pmatrix}
|\varphi^{+}\rangle \\
0%
\end{pmatrix}
$ and $\sigma_{z}
\begin{pmatrix}
0 \\
|\varphi^{-}\rangle
\end{pmatrix}
= -
\begin{pmatrix}
0 \\
|\varphi^{-}\rangle
\end{pmatrix}
$. Meanwhile, $\sigma_{z}= U \Pi U^{+}$ is actually the parity
operator of $H_{SB}$ after the unitary transformation $U$ has been
performed, which means that $H_{SB}$ also consists of two
sub-Hamiltonians with the eigenfunctions of even-parity and
odd-parity, respectively, with respect to the parity operator $\Pi$.

If QPT happens in the model of $H_{SB}$, there should be the
possibility of degenerate ground states of the two decoupled
sub-Hamiltonians $H^{+}$ and $H^{-}$ and also the possibility of
releasing the degeneracy \cite {hur1,Bu1,vojta1,hur,winter}. To
check the possibilities, we may solve $H^+$ and $H^{-}$
collaboratively by expanding $\ket{\varphi^+}$ and $\ket{\varphi^-}$
using displaced coherent states
\begin{equation}
\label{eq:phi}
\begin{split}
\ket{\varphi^+} &= \textstyle{\sum_{\{n\}}} c^+_{\{n\}} \ket{\{ n
\}}_A, \\ \ket{\varphi^-} &= \textstyle{\sum_{\{n\}}} c^-_{\{n\}}
e^{i\pi\sum_{k}a_{k}^{\dagger}a_{k}} \ket{\{ n \}}_A,
\end{split}
\end{equation}
where $c_{\{n\}}^{\pm}$ are coefficients to be determined, $\{n\}=n_{1}, \cdots, n_{N}$ are for
different bosonic modes, $\ket{\{ n \}}_A = \prod^N_{k=1} \frac
{e^{-q_{k}^{2}/2}}{\sqrt{n_{k}!}}(a^{\dagger}_{k}+q_{k})^{n_{k}}e^{-q_{k}a_{k}^{\dagger}}\ket{0}$
is the product of displaced coherent states of different modes of
the bosonic field with the displacement variables
$q_{k}=\lambda_{k}/\omega_{k}$, $k=1, 2, \cdots, N$, and $\ket{0}$
the vacuum state of the bosons.

Expanding the tunneling terms in Eq. \eqref{eq:H+-} by $\{ \ket{\{ n \}}_A \}$ yields
\begin{equation}
\label{eq:expanding} \frac{\Delta}{2} e^{i \pi \sum_k
a^\dagger_k a_k} =  \frac{\Delta}{2}\sum_{\{ m \},\{ n \}}\!\!
D_{\{ m,n \}}\ket{\{ m \}}_{AA} \!\!\bra{\{ n \}},
\end{equation}
where $D_{\{m,n\}}$ is given by \cite {TLiu}
\begin{equation}
\label{eq:Dmn}
D_{\{m,n\}} = e^{-2\sum_{k}q_{k}^{2}}L_{\{ m,n \}}
\end{equation}
with $$L _{\{ m,n \}} = \prod_{k=1}^{N}\sum_{j=0}^{\min
[m_k,n_k]}(-1)^{j}\frac{\sqrt{m_{k}!n_{k}!}(2q_{k})^{m_{k}+n_{k}-2j}}{(m_{k}-j)!(n_{k}-j)!j!}.$$
So the prefactor $e^{-2\sum_k q_k^2}$, relevant to $D_{\{m,n\}}$ and
the tunneling terms, plays a crucial role in the SBM and should be
non-zero. Otherwise the SBM in our case is physically meaningless due to neither
local field nor tunneling involved. This point will be further
emphasized later.

The solutions of $H^+$ and $H^{-}$ by Eq. (6) yield
\begin{equation}
E^{\pm}_{\{m\}} = \sum_k \omega_k (m_k - q_k^2) \mp \frac{\Delta}{2
c^{\pm}_{\{m\}}} \sum_{\{n\}} c^{\pm}_{\{n\}} D_{\{m,n\}}.
\end{equation}
Thus the lowest eigenenergies for the even- and odd-parity states
are, respectively,
\begin{equation}
\label{eq:E0}
E^{\pm}_{\{0\}} = -\sum_k \omega_k q_k^2 \mp \frac {\Delta}{2c^{\pm}_{\{0\}}}
\sum_{\{n\}} c^{\pm}_{\{n\}} D_{\{0,n\}}.
\end{equation}
It is intuitive to reach degeneracy by setting $\Delta=0$, which is,
however, physically meaningless in our case. For any non-vanishing
value of $\Delta$, we show a rigorous proof in Appendix that there
is no possibility of $E_{\{0\}}^+ = E_{\{0\}}^-$. That is to say, a
physically meaningful solution of the SBM in our case requires no
degeneracy for the ground states of $|\varphi ^{+}\rangle$ and
$|\varphi^{-}\rangle$, which implies no QPT happening, in terms of
the viewpoints in \cite {hur1,Bu1,kehrein,vojta1,hur,winter}, no
matter how to change the characteristic parameters and which bath is
considered.

Our results also mean that the ground state of $H_{SB}$ is surely of
a certain parity, and any employed state with mixed parity (e.g.,
the superposition of the ground states of $|\varphi ^{+}\rangle$ and
$|\varphi^{-}\rangle$) is definitely not the ground state of the
model. In this sense, it is not a surprise that QPT was found in
\cite {chin} using a presumed ground state which, due to mixture of
odd- and even-parity, is actually not for the ground state of the
model. In fact, our result could be obtained more simply by
calculating the magnetization $M=\langle\sigma_{z}\rangle$. Without
losing generality, we assume the eigenfunction of $H_{d}$ with the
two sub-eigensolutions correlated, i.e., $|\Psi\rangle=
\begin{pmatrix}
\cos\theta |\varphi^{+}_{0}\rangle \\
\sin\theta |\varphi^{-}_{0}\rangle%
\end{pmatrix}
$, where $|\varphi^{\pm}_{0}\rangle$ are the ground states,
respectively, of $|\varphi^{\pm}\rangle$. So the magnetization of
the Hamiltonian $H_{SB}$ is
\begin{equation}
M=\langle\sigma_{z}\rangle = \langle\Psi| U \sigma_{z} U^{+} |\Psi\rangle =
-\sin 2\theta\langle\varphi^{+}_{0}|e^{i\pi\sum_{k}a^{\dagger}_{k}a_{k}} |
\varphi^{-}_{0}\rangle.
\end{equation}
If $\theta=k\pi$ or $(k+1/2)\pi$ with $k=0, 1, \cdots$, which
corresponds to even-parity case or odd-parity case, the
magnetization $M$ is absolutely zero, which implies no QPT
happening. The finite value of $M$ occurs only in the case of the
superposition of $|\varphi ^{+}_{0}\rangle$ and
$|\varphi^{-}_{0}\rangle$, which, according to our discussion above,
is not the ground state of the model.

Besides the incorrect employment of the state as the ground state
\cite {chin}, the main reason for obtaining the QPT in previous
literatures \cite {Bu1,hur1,kehrein,vojta2,vojta1, hur,
winter,Bu2,Bu3,and,rmp1,cheng,
AA,chin,vojta3,wong,florens,zheng1,vojta4,zheng2} is the
unreasonable treatment of the infrared divergence existing in  the
spectral functions for Ohmic and sub-Ohmic dissipations, which
induced the degeneracy of the ground states in the low frequency
domain. To clarify this point, we demonstrate below the potential
infrared divergence in the treatments using the bath mode with
continuous and discretized spectra, respectively. The NRG is taken
as an example to show that the prediction of the QPT in previous
studies is completely due to divergent expansion in the infrared
limit.

\textit{For treatment with the bath mode of continuous spectrum: }
We return to the prefactor $e^{-2\sum_k q_k^2}$, for which, using
the spectral function $J(\omega)$, we have,
\begin{eqnarray}
\sum_{k}q_{k}^{2}=\sum_{k}\lambda_{k}^{2}/\omega^{2}_{k} =\int_{0}^{\omega_{c}}
\sum_{k}\frac {\lambda_{k}^{2}}{\omega^{2}}\delta(\omega-\omega_{k})d\omega  \notag \\
=\int_{0}^{\omega_{c}}2\alpha\omega_{c}^{1-s}\omega^{s-2}d\omega =
2\alpha\beta_{1}, ~~~~~~~~~~~~~
\end{eqnarray}
with
\begin{displaymath}
\beta_{1}=\left\{ \begin{array}{ll} \frac{1}{s-1} [1-(\frac{\omega_c}{\omega_1})^{1-s}] & \textrm{if $s<1$} \\
\ln (\frac{\omega _{c}}{\omega_{1}})  & \textrm{if $s=1$} \\
\frac{1}{s-1} & \textrm{if $s>1$}, \end{array} \right.
\end{displaymath}
where $\alpha$ and $\omega_{c}$ are defined above in the spectral
function.  $\omega_{1}$ is a small quantity regarding the frequency
difference from $\omega=0$. In the case of the infrared limit, i.e.,
$\omega_{1}\rightarrow 0$, we have
$\sum_{k}q_{k}^{2}\rightarrow\infty$ if $s\leqslant 1$,  which leads
to $D_{\{m,n\}}=0$ (See Eq. (8)) and thereby
$E_{\{0\}}^{+}=E_{\{0\}}^{-}$ from Eq. (10). As a result, with
calculation approaching the low frequency domain, the NRG gives a
significant change from non-degeneracy to degeneracy, which was
called QPT. In addition, we see from Eq. (7) that the tunneling
terms are vanishing with the calculation approaching the infrared
limit, which reduces $H_{SB}$ to be
\begin{equation*}
H'=\sum_{k}\omega_{k}a^{\dagger}_{k}a_{k} +
\sum_{k}\lambda_{k}(a^{\dagger}_{k} + a_{k})\sigma_{z},
\end{equation*}
commuting with $\sigma_{z}$. This corresponds to the localization,
sometimes also called the frozen spin \cite{hur}. It means that,
once the system falls into a well-polarized state of the spin, the
magnetization will remain unchanged in the dynamics. Consequently,
it is the infrared divergence that wrongly predicts the QPT
happening from the delocalization to the localization as the
infrared limit is approached.

Eq. (12) presents the relationship between the prefactor and the
dissipation strength $\alpha$, from which we see no possibility to
have a finite value of $\alpha$ as the critical number for the QPT.
The degeneracy of the ground states occurs only in the case of
$\alpha\rightarrow\infty$, i.e., a physically meaningless case. On
the other hand, considering Eq. (7), we may find that the infrared
divergence also yields the parity operator
$\Pi=\sigma_{x}e^{i\pi\sum_{k}a^{\dagger}_{k}a_{k}}$ to be zero. In
fact, the correct understanding of the parity variation could be
reached only after we have eliminated the influence from the
infrared divergence. In such a case, the parity of the ground state
of the model will explicitly break down as long as we introduce the
non-zero local field into $H_{SB}$, which brings about a significant change in the
magnetization as a function of $\epsilon$. We prefer to call this
phenomenon as parity breaking, instead of the QPT.

\begin{figure}[htbp]\label{Fig1}
\includegraphics[width=2.7in]{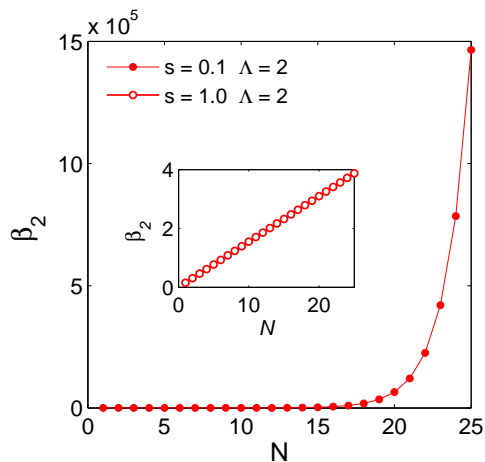}
\caption{(Color online) Divergence of $\beta_{2}$ in the calculation by
NRG, where $\Lambda =2$, $\omega_c =1$, and $s=0.1$ ($s=1.0$ in
inset). }
\end{figure}

\textit{For treatment with the bath mode of discretized spectrum: }
By means of the NRG logarithmic discretization parameter $\Lambda$
defined in \cite{Bu1,vojta1}, we have $\sum_k q_k^2 = 2 \alpha
\beta_{2}$, where
\begin{displaymath}
\beta_{2}=\left\{ \begin{array}{ll} \frac{1}{4}\beta_0 \frac{\Lambda^{(1-s)(N+1)} -1 }{\Lambda^{1-s} -1} &s \neq 1 \\
\frac{1}{4}\beta_0 (N+1) & s=1, \end{array} \right.
\end{displaymath}
with $N$ a large number required by the method \cite{Bu1},
$\beta_{0}=(s+2)^{2}(1-\Lambda^{-s-1})^{3}/[(s+1)^{3}(1-\Lambda
^{-s-2})^{2}]$.  So $\sum_{k}q_{k}^{2}\rightarrow\infty$ in the case
of $N\rightarrow\infty$ and $s\leqslant 1$. As an example, we show
in Fig. 1 the distribution of $\beta_{2}$ with respect to
the mode-number $N$ for different $s$. The divergence leads to
$E_{\{0\}}^{+}=E_{\{0\}}^{-}$ in the case of Ohmic and sub-Ohmic
dissipations, from which the QPT was predicted in previous studies.

Our work not only indicates for the first time that the parity
keeping and breaking, rather the QPT, can fully characterize the
behavior in the SBM, but also helps to understand a fact that the
QPT had only been found by previous numerics in the sub-Ohmic and
the Ohmic cases, instead of in the super-Ohmic case, which is
because the infrared divergence never happens in the super-Ohmic
case, e.g, $\sum_{k}q_{k}^{2}=2\alpha/(s-1)$ for $s>1$ in the case
of the continuous bath-mode spectrum. In fact, our investigation
above has clearly shown that, whatever the bath-mode spectrum is, no
QPT would happen if the local field is absent in the SBM.

In summary, we have presented analytical treatments to show reliably
the impossibility of the QPT in the SBM without the local field.
Since our treatments are independent of the bath type, the degrees
of freedom of the bath and the calculation precision, we argue that
the previous conclusions drawn for the QPT happening in the Ohmic
and sub-Ohmic SBM need more serious reexamination. The parity
discovered in the present work is strongly relevant to the
configuration of the model and the parity breaking can significantly
change the dynamics of the model. In addition, the SBM has wide
application ranging from the electron transfer in biomolecules \cite
{bio} to the entanglement in quantum information science
\cite{weiss,leggett}. Therefor, we believe that our study is of
general interest and would be helpful for our deeper understanding
of the weird phenomena in open quantum systems.

This work is supported by National Fundamental Research Program of
China (Grant No. 2012CB922102), by National Natural Science
Foundation of China under Grants No. 10974225 and No. 11004226, and
by funding from WIPM.
\\

\section*{Appendix: Proof of the impossibility of degeneracy}

The proof starts from Eq. (10). Excluding the trivial case $\Delta
=0$, we have $E^+_{\{0\}} = E^-_{\{0\}}$ occurring only for (1)
$\sum_{\{n\}} c^{\pm}_{\{ n \}} D_{\{ 0,n \}} = 0$ or (2) $\sum_{\{
n \}} c^{+}_{\{ n \}} D_{\{ 0,n \}} / c^+_{\{ 0 \}} = -\sum_{\{ n
\}} c^{-}_{\{ n \}} D_{\{ 0,n \}} / c^-_{\{ 0 \}} \neq 0$.

Employing the result
\begin{equation*}
D_{\{0 ,n\}}=e^{-2\sum_k q_k^2}\prod_k \frac {(2 q_k)^{n_k}}{\sqrt{n_k!}},
\end{equation*}
we first check the case (1). Since we require $e^{-2 \sum_k q_k^2}
\neq 0$, the case $\sum_{\{ n \}} c^{\pm}_{\{ n \}} D_{\{ 0,n \}} =
0$ implies $\sum_{\{ n \}} c^\pm_{\{ n \}} \prod_k \left[ (2
q_k)^{n_k} /\sqrt{n_k!} \right] =0$, which is not satisfied unless
$c^\pm_{\{ n \}} = 0$. So the case (1) makes no sense and can be
dropped.

For the case (2), if we have  $\sum_{\{ n \}} c^{+}_{\{ n \}} D_{\{
0,n \}} / c^+_{\{ 0 \}} = -\sum_{\{ n \}} c^{-}_{\{ n \}} D_{\{ 0,n
\}} / c^-_{\{ 0 \}} \neq 0$, which leads to $E^{+}_{\{0\}} =
E^{-}_{\{0\}}$, then we reach
\begin{equation*}
2D_{\{0,0\}}+\frac {1}{c^+_{\{0\}}}\sum_{\{n\}\neq 0} c^+_{\{n\}} D_{\{
0,n\}}= \frac {-1}{c^-_{\{0\}}}\sum_{\{n\}\neq 0} c^-_{\{n\}} D_{\{0,n\}},
\end{equation*}
where $\sum_{\{n\}\neq 0}$ is the summation excluding the vacuum state of each bosonic mode. Since
both sides of the above equation are of multivariate polynomials of $q_{k}$,
we may compare the terms of $q_{k}$ to determine if such an equation is
satisfied. To this end, we further deduce the equation as
\begin{equation*}
2+ \frac {1}{c^+_{\{0\}}}\sum_{\{n\}\neq 0}c^-_{\{n\}}\prod_k  \frac{(2 q_k)^{n_k}}{\sqrt{n_k!}}
= \frac {-1}{c^-_{\{0\}}}\sum_{\{n\}\neq 0} c^+_{\{n\}}\prod_k \frac{(2 q_k)^{n_k}}{\sqrt{n_k!}}.
\end{equation*}
It is evident that the equation could not be satisfied since no constant term exists
in the right side of the equator. This is contradictory to the above presumption
$E^+_{\{0\}}=E^-_{\{0\}}$. So we have proven the impossibility of degeneracy
in the case of $\Delta\neq 0$.

\end{document}